\documentclass[doublecol]{epl2}

\usepackage{graphicx}
\usepackage{subfigure}
\usepackage{amsthm}
\usepackage{amsmath}
\usepackage{amssymb}
\usepackage{verbatim}
\usepackage{dcolumn}
\usepackage{bm}
\usepackage{epsf}
\usepackage{color}
\usepackage[colorlinks=true,citecolor=blue,linkcolor=blue,urlcolor=blue]{hyperref}%
\usepackage{xcolor}
\usepackage{dsfont}
\newcommand{\bra}[1]{\left\langle #1\right|}
\newcommand{\ket}[1]{\left|#1\right\rangle}
\newcommand{\braket}[2]{\left\langle #1|#2\right\rangle}

\newcommand{\tr}[1]{\mathrm{tr}\left\{#1\right\}}

\newcommand{\pd}{\partial}

\newcommand{\id}{\mathbb{I}}
\newcommand{\com}[2]{\left[#1,\,#2\right]}

\newcommand{\bla}{bla\\bla\\bla\\bla\\bla}

\newcommand{\mc}[1]{\mathcal{#1}}

\newcommand{\mrm}[1]{\mathrm{#1}}

\title{Suppressing excitations in the nonlinear Landau-Zener model}

\author{Sebastian Deffner\inst{1,2,3} \thanks{E-mail: \email{deffner@umbc.edu}} \and Steve Campbell\inst{4,5} \thanks{E-mail: \email{steve.campbell@ucd.ie}} }
\shortauthor{Sebastian Deffner and Steve Campbell}

\institute{\inst{1} Department of Physics, University of Maryland, Baltimore County, Baltimore, MD 21250, USA\\
\inst{2} Quantum Science Institute, University of Maryland, Baltimore County, Baltimore, MD 21250, USA\\
\inst{3} National Quantum Laboratory, College Park, MD 20740, USA\\
\inst{4}School of Physics, University College Dublin, Belfield, Dublin 4, Ireland\\
\inst{5}Centre for Quantum Engineering, Science, and Technology, University College Dublin, Dublin 4, Ireland}

\date{\today}

\abstract{Many complex quantum systems can be described by effectively nonlinear dynamics. While such dynamics have many appealing characteristics, they also make the analysis significantly more involved. This is due to the fact that only a few analytical treatments exist, and that the language of quantum mechanics is built for linear operators. For instance, the very formulation of the quantum adiabatic theorem requires the underlying dynamics to be linear. In this work we show that in a generalized Landau-Zener model, nonlinear dynamics can be leveraged to suppress excitations and coherences of the corresponding linear scenario. To this end, we introduce a generalized ``energy spectrum'', which is defined by the expectation values of the energy under the stationary states. As a main result, we show that the nonlinear term in the evolution equation acts like an effective shortcut to adiabaticity for the linear Landau-Zener problem.}

\begin{document}

\maketitle

\section{Introduction}

If one accepts that any quantum system can co-exist with another without interaction, it can be shown that all quantum dynamics must fundamentally be linear \cite{Jordan2009JPCF}. However, in  complex quantum many-body systems, the properties of collective excitations can often be effectively described by non-linear Schr\"odinger equations \cite{DalFavero2024PRA}. Among these, the Gross-Pitaevskii equation \cite{Gross1961,Pitaevskii1961} is probably most widely known, with applications ranging from Bose-Einstein condensation \cite{Gross1961,Pitaevskii1961} to nonlinear optics \cite{Rand2010} to plasma physics \cite{Ruderman2002}.

From the point of view of quantum information processing, nonlinear variants of quantum mechanics are very appealing as the corresponding dynamics can facilitate processes that are not possible in linear systems, such as, e.g., perfectly distinguishing non-orthogonal states \cite{Childs2016PRA} or synchronization \cite{Zhang2020MPLB,Shen2023PRA}. It is, thus, not a surprise that significant work has been dedicated to the development of ``nonlinear quantum computation'' \cite{Meyer2014PRA,Lacy2018QIP,Chiew2019QIP,Holmes2023PRA,Geller2023CTP,Geller2023AQT,Deiml2024arXiv}. The effectively nonlinear dynamics are then generated by, for instance, coupling Bose-Einstein condensates \cite{Byrnes2015,Xu2022PRR,Geller2024AQT,Grossardt2024arXiv}, building qubits from chaotic orbits \cite{Geller2023SR}, intermittent measurements \cite{Kalman2018PRA,Sakaguchi2023NC}, or nonlinear optics \cite{Yang2008CTP,Chang2014,Gu2016AOP,Scala2024CP}.

In the present letter, we report a comprehensive analysis of the Landau-Zener model \cite{landau1932theorie,zener1932non,stuckelberg1932theorie} undergoing general, nonlinear quantum dynamics. This model was recently proposed in the context of nonlinear quantum thermometry \cite{Deffner2025QST}, where it was found that the nonlinear quantum dynamics can be leveraged to dramatically enhance the precision of measurements. 

More broadly, the Landau-Zener model is probably one of the best studied quantum systems, as its dynamics are essentially analytically solvable \cite{Barnes2012PRL,Barnes2013PRA,Glasbrenner2023JPB}, and it provides a simple and tractable model for quantum phase transitions \cite{Damski2005PRL}. Thus, the Landau-Zener model is also the ``go-to system'' for pedagogical and accessible studies of nonequilibrium quantum processes, and how, e.g., to suppress excitations arising from finite-time driving \cite{Puebla2020PRR,Menu2022, Soriani2022PRA,Cavalcante2025arXiv, Sveistrys2025}. For instance, the Landau-Zener model is also one of the few systems for which the counterdiabatic field can be written in closed form \cite{Berry2009JPA,Campbell2017PRL}.

Building on the insights obtained in Ref.~\cite{Deffner2025QST}, one is tempted to inquire whether nonlinear dynamics can also be leveraged to suppress (or enhance) parasitic excitations arising from finite-time driving. However, this is not necessarily a well-posed question, as ``excitations'' usually refers to transitions between the eigenstates of the Hamiltonian. In quantum mechanics eigenstates, as well as eigenvalues, are usually reserved for linear maps. While mathematically, nonlinear eigenvalue problems can be well-defined \cite{Chiappinelli2018axioms}, it is far from obvious how the quantum adiabatic theorem would generalize to the corresponding nonlinear dynamics. 

Thus, in our analysis, we take a pragmatic approach. We initialize the model in stationary state, and study the properties of the quantum state evolved under the nonlinear Schr\"odinger equation. Interestingly, we find that the nonlinear Landau-Zener model has two stationary states, which correspond to lower and higher expected energies. ``Excitations'' are then understood by computing the overlap of the time-evolved states with the instantaneous stationary states. Remarkably, we find that such excitations are strongly suppressed for nonlinear dynamics. Moreover, we show that also coherences in the energy basis are suppressed. Thus we can conclude that, in a certain sense, nonlinear dynamics act akin to (approximate) shortcuts to adiabaticity for the corresponding linear dynamics, in line with other recent studies~\cite{Li2018}.

\section{Nonlinear Landau-Zener model}

We start by establishing notions and notations for the Landau-Zener model and its nonlinear dynamics. In the following, we consider general quantum dynamics described by the nonlinear Schr\"odinger equation \cite{Childs2016PRA}
\begin{equation}
\label{eq:nonlin_sch}
i \, \pd_t \ket{\psi_t}=H_t\,\ket{\psi_t}+ K \ket{\psi_t}
\end{equation}
where $H_t$ is the usual, Hermitian, time-dependent Hamiltonian. To avoid clutter in the formulas, we will work in units for which $\hbar=1$. Further, $K$ describes a general nonlinearity of the form
\begin{equation}
\bra{x} \left(K \ket{\psi_t}\right)= \mc{K}(|\braket{x}{\psi_t}|)\,\braket{x}{\psi_t}\,.
\end{equation}
Such dynamics have been extensively studied recently, in e.g. state discrimination \cite{Childs2016PRA}, quantum speed limits \cite{Deffner2022EPL}, and quantum thermometry \cite{Deffner2025QST}. However, almost all nonlinear quantum dynamics of relevance can be written in this form. For instance, for the Gross-Pitaeveskii equation \cite{Gross1961,Pitaevskii1961} we simply have $\mc{K}(|\braket{x}{\psi_t}|)=\kappa\,|\braket{x}{\psi_t}|^2$, and for the Kolomeisky equation \cite{Kolomeisky2000PRL}, $\mc{K}(|\braket{x}{\psi_t}|)=\kappa\,|\braket{x}{\psi_t}|^4$. Even more intricate dynamics have been considered for Bose liquids \cite{Meyer2014PRA}, in which case we have $\mc{K}(|\braket{x}{\psi_t}|)=\kappa\, \log|\braket{x}{\psi_t}|^2$.

\subsection{Nonlinear Landau-Zener dynamics}

We now consider a nonlinear generalization of the Landau-Zener model \cite{landau1932theorie,zener1932non,stuckelberg1932theorie}. Its Hamiltonian can be written as \cite{Soriani2022PRA}
\begin{equation}
\label{eq:H_LZ}
H_t=\Delta_t\, \sigma_z+ J \sigma_x\,
\end{equation}
where $\Delta_t$ is the time-dependent magnetic field and $J$ is the coupling strength.

In the following it will be convenient to write the quantum state $\ket{\psi_t}$ in its Bloch representation,
\begin{equation}
\label{eq:bloch}
    \rho_t=\ket{\psi_t}\bra{\psi_t}=\frac{1}{2}\left(\id_2+\vec{r}_t\cdot\vec{\sigma}\right)\,,
\end{equation}
where $\vec{r}_t=(x_t,y_t,z_t)$ is the Bloch vector, and $\vec{\sigma}=(\sigma_x,\sigma_y,\sigma_z)$ is the Pauli vector. Note that for pure states $|\vec{r}|=1$. It is also worth emphasizing that the nonlinear Schr\"odinger equation \eqref{eq:nonlin_sch} is purity preserving, i.e., if $|\vec{r}_0|=1$ then $|\vec{r}_t|=1$ for all $t$. Note that the corresponding propagator is \emph{not} unitary, since the dynamics is \emph{not} linear.

In the Bloch representation \label{eq:bloch}, the non-linear term, $\mc{K}_t$, drastically simplifies. Since $|\braket{0}{\psi_t}|^2=(1+z_t)/2$ and $|\braket{1}{\psi_t}|^2=(1-z_t)/2$, $\mc{K}_t$ can be written as an effectively state-dependent Hamiltonian \cite{Childs2016PRA},
\begin{equation}
\mc{K}_t=\begin{pmatrix}
\mc{K}\left(\sqrt{(1+z_t)/2}\right)&0\\
0&\mc{K}\left(\sqrt{(1-z_t)/2}\right)\,.
\end{pmatrix}
\end{equation}
Hence, the non-linear Schr\"odinger equation \eqref{eq:nonlin_sch} is equivalent to,
\begin{equation}
\label{eq:diffeq_LZ}
\begin{aligned}[b]
\dot{x}_t&=-\left[2\Delta_t +\tilde{\kappa}(z_t)\right]\,y_t\\
\dot{y}_t&=\quad\left[2\Delta_t +\tilde{\kappa}(z_t)\right]x_t-2 J z_t\\
\dot{z}_t&=\quad 2 J y_t\,.
\end{aligned}
\end{equation}
where $\tilde{\kappa}(z_t)\equiv\mc{K}(\sqrt{(1+z_t)/2})-\mc{K}(\sqrt{(1-z_t)/2})$.
 
\subsection{Stationary states and expected energies}

\begin{figure*}
    \centering
\includegraphics[width=0.3\linewidth]{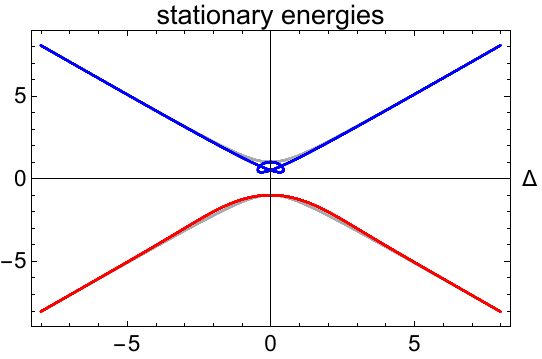}\hfill\includegraphics[width=0.3\linewidth]{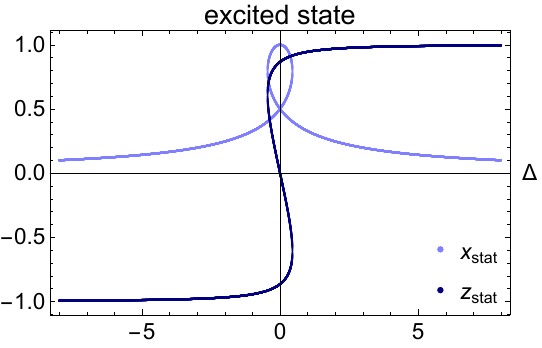}\hfill \includegraphics[width=0.3\linewidth]{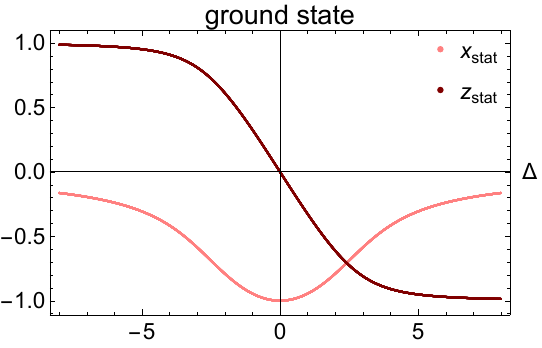}\\
    \caption{(left panel) Generalized energy ``spectrum'' for the nonlinear Landau-Zener model with Gross-Pitaevskii nonlinearity, $\tilde{\kappa}(z)=\kappa\, z$. Gray lines correspond to the spectrum of the linear Landau-Zener model. Stationary Bloch vector of the ``excited'' state (middle panel) and ``ground'' state (right panel). Parameters are $J=1$ and $\kappa=4$.}
    \label{fig:stat}
\end{figure*}

Before analyzing the time-dependent dynamics, we need to consider the stationary solutions. It is once again important to recognize that in quantum mechanics familiar concepts, such as eigenspectra and eigenstates are exclusively associated with linear operators. Hence, the non-linear Landau-Zener model cannot be described in these familiar terms.

However, it is easy to see that Eq.~\eqref{eq:diffeq_LZ} has a well-defined stationary solution for any value $\Delta_t=\Delta$. Recognizing $y_\mrm{stat}=0$, the stationary Bloch vector is given by the solutions of the algebraic equations
\begin{equation}
\begin{split}
\label{eq:stat}
0=\left[2\Delta +\tilde{\kappa}(z_\mrm{stat})\right]x_\mrm{stat}\quad \text{and}\quad 1=x_\mrm{stat}^2+z^2_\mrm{stat}\,,
\end{split}
\end{equation}
where the second equation guarantees the purity of the stationary states. The generalized ``spectrum'' is then given by the expected energies corresponding to the solutions of \eqref{eq:stat},
\begin{equation}
E_\mrm{stat}=\tr{H\,\rho_\mrm{stat}}=\Delta z_\mrm{stat}+J x_\mrm{stat}\,.
\end{equation}

\begin{figure*}
    \centering
\includegraphics[width=0.3\linewidth]{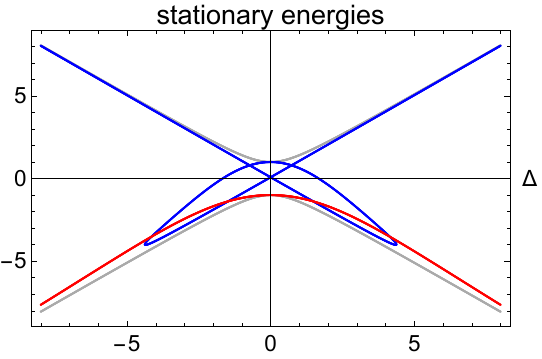}\hfill\includegraphics[width=0.3\linewidth]{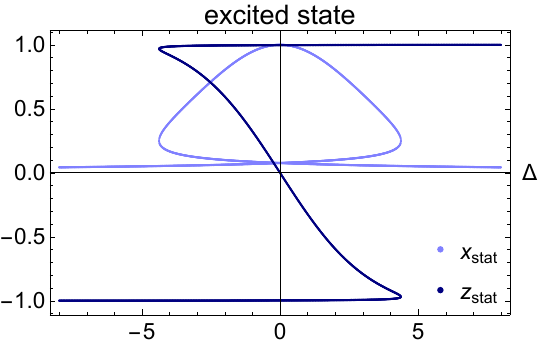}\hfill \includegraphics[width=0.3\linewidth]{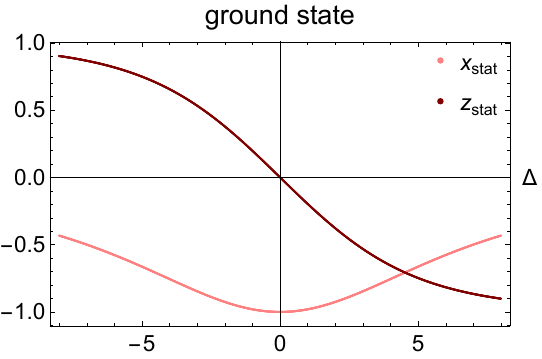}\\
    \caption{(left panel) Generalized energy ``spectrum'' for the nonlinear Landau-Zener model for Bose liquids with a logarithmic term, $\tilde{\kappa}(z)=\kappa\, \mrm{arctan}(z)$.  Gray lines correspond to the spectrum of the linear Landau-Zener model. Stationary Bloch vector of the ``excited'' state (middle panel) and ``ground'' state (right panel). Parameters are $J=1$ and $\kappa=4$.}
    \label{fig:stat_log}
\end{figure*}

In Fig.~\ref{fig:stat} we plot the resulting stationary energies for Gross-Pitaevskii nonlinearities with $\tilde{\kappa}(z)=\kappa\, z$, and in Fig.~\ref{fig:stat_log} for Bose liquids, in which case we have $\tilde{\kappa}(z)=\kappa\, \mrm{arctan}(z)$.

In both cases, we observe that the ``ground'' state is remarkably similar to the linear case. However, the energy of the ``excited'' state exhibits a characteristic ``pretzel'' around $\Delta=0$, which arises from Eq.~\eqref{eq:stat} having multiple real roots. Moreover, we find that these nonlinear effects are more pronounced for Bose liquids, i.e., for logarithmic nonlinearities. Also note that since the stationary states are \emph{not} eigenstates of the Hamiltonian, crossings in the energy diagram are not indicative of certain transitions.

The obvious question arises to what extent these nonlinear modifications of the energy ``spectrum'' affect transitions between ``ground'' and ``excited'' in finite-time processes. To this end, we now continue the analysis for the usual Landau-Zener problem.

\section{Suppression of excitations and coherences}

In the original Landau-Zener problem \cite{landau1932theorie,zener1932non,stuckelberg1932theorie} one is interested in determining the probability of transitions from ground to excited states, if the system is driven through the avoided crossing at finite, constant rate. 

Thus, we now consider the nonlinear dynamics \eqref{eq:diffeq_LZ} for $\Delta_t=\alpha t/\tau$ and $-\tau/2\leq t \leq \tau/2$. In the linear case, $\kappa=0$, the Landau-Zener formula for such protocols can be written as \cite{Soriani2022PRA,Soriani2022PRA2}
\begin{equation}
\label{eq:LZ}
\mc{P}_\tau^{\kappa=0}=\frac{1}{2}\,\left(1+z_{\tau/2}^\mrm{\kappa=0}\right)=\exp{\left(-\pi J^2 \tau/\alpha\right)}\,,
\end{equation}
which describes the exact dynamics to very high accuracy \cite{Soriani2022PRA,Soriani2022PRA2,Cavalcante2025arXiv}.

\subsection{Time-dependent dynamics}

\begin{figure}
    \centering
    \includegraphics[width=0.48\textwidth]{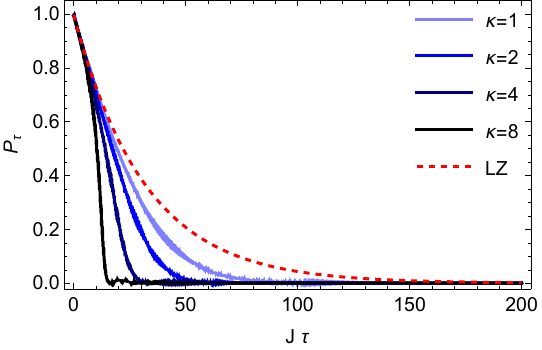}
    \caption{Probability to find the nonlinear Landau-Zener model in the excited state, $\mc{P}_\tau=(1+z_{\tau/2})/2$, for a Gross-Pitaeveskii nonlinearity, $\tilde{\kappa}(z)=\kappa\, z$, and $\alpha=100$. Red, dashed line depicts the Landau-Zener formula Eq.~\eqref{eq:LZ}.}
    \label{fig:prob_x2}
\end{figure}

\begin{figure}
    \centering
    \includegraphics[width=0.48\textwidth]{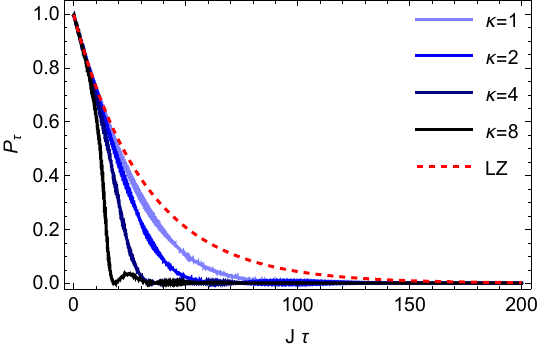}
    \caption{Probability to find the nonlinear Landau-Zener model in the excited state, $\mc{P}_\tau=(1+z_{\tau/2})/2$, for a logarithmic nonlinearity,  $\tilde{\kappa}(z)=\kappa\, \mrm{arctan}(z)$, and $\alpha=100$. Red, dashed line depicts the Landau-Zener formula Eq.~\eqref{eq:LZ}.}
    \label{fig:prob_log}
\end{figure}

Generally, the nonlinear dynamics \eqref{eq:diffeq_LZ} are not analytically solvable. Hence, we have to rely on a numerical analysis. To this end, we solve Eq.~\eqref{eq:diffeq_LZ} for the linear protocol, $\Delta_t=\alpha t/\tau$ with $-\tau/2\leq t \leq \tau/2$ and $\alpha=100$. As initial state we chose the ``ground'' state, and we compute the probability to find the two-level system in its excited state at $t=\tau/2$. In Fig.~\ref{fig:prob_x2} we depict the resulting transition probability $\mc{P}_\tau=(1+z_{\tau/2})/2$ for the Gross-Pitaevskii dynamics and in Fig.~\ref{fig:prob_log} for Bose liquids.

Despite the marked difference of the ``pretzel'' in the generalized spectra, cf. Figs.~\ref{fig:stat} and \ref{fig:stat_log}, the resulting transition probabilities are remarkably similar. Thus it will suffice to continue the analysis for only one of the cases, and without loss of generality we choose the Gross-Pitaevskii scenario.

For very fast processes as well as for very slow processes, the Landau-Zener formula \eqref{eq:LZ} correctly captures the dynamical behavior. However, for intermediate rates between these extremes we observe that a stronger nonlinearity results in the transition probability approaching zero earlier. In other words, the nonlinear term in Eq.~\eqref{eq:diffeq_LZ} effectively suppresses transitions between the instantaneous stationary states. 

\begin{figure}
    \centering
    \includegraphics[width=0.48\textwidth]{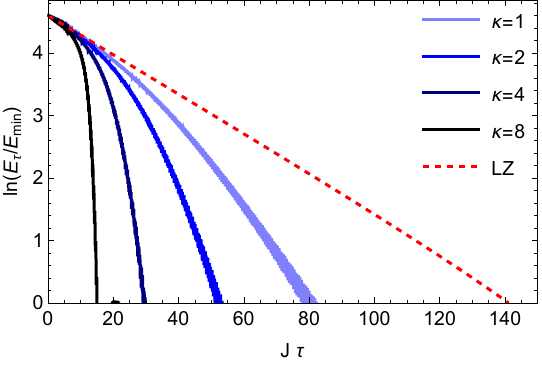}
    \caption{Expected energy \eqref{eq:energy} normalized by the minimal ground state energy (obtained for $\tau\gg1$) at $t=\tau/2$ for a Gross-Pitaeveskii nonlinearity, $\tilde{\kappa}(z)=\kappa\, z$, and $\alpha=100$.}
    \label{fig:energy_x2}
\end{figure}

This is further corroborated by inspecting Fig.~\ref{fig:energy_x2}, in which we plot the final expected energy,
\begin{equation}
\label{eq:energy}
  E_{\tau}=\tr{H_{\tau/2}\rho_{\tau/2}} = \frac{\alpha}{2}\, z_{\tau/2}+J\, x_{\tau/2},
\end{equation} 
for Gross-Pitaevskii dynamics. We observe that the stronger the nonlinearity, the faster (i.e. smaller $\tau$) we find the final energy is identical to the ``ground'' state energy. 

Since for $\alpha\gg1$ the stationary states of the nonlinear model become virtually identical to the energy eigenstates of the bare Landau-Zener Hamiltonian, one might interpret the nonlinear term as some kind of effective shortcut to adiabaticity \cite{Guery2019RMP,Hatomura2024JPB}. However, such a conclusion needs to be analyzed more carefully since the quantum adiabatic theorem does not apply to nonlinear dynamics, and \emph{a priori} it cannot be ruled out that the system builds up appreciable quantum coherence between the two stationary states.

\subsection{Suppression of coherence}

Quantum coherence can be understood as a resource available to genuine quantum systems \cite{Baumgratz2014PRL}. Generally, the term coherence refers to the degree to which quantum states overlap with a chosen basis. Therefore, the amount of quantum coherence depends on the representation of quantum states. 

An often studied measure quantifying the amount of quantum coherences supported by a quantum state is the \emph{relative entropy of coherence} \cite{Bu2017PRL,Gao2019IJTP,Song2020IJTP,Huang2023Photonics,Lecamwasam2024PRXQ}, which in the Bloch representation reads \cite{Baumgratz2014PRL}
\begin{equation}
\label{eq:coh}
\begin{split}
&\mc{C}(\rho)=S(\rho^\mrm{diag})-S(\rho)\\
&\quad=-\frac{1}{2}(1-z)\ln\left[\frac{1}{2}(1-z)\right]-\frac{1}{2}(1+z)\ln\left[\frac{1}{2}(1+z)\right]\,.
\end{split}
\end{equation}
Here, $\rho^\mrm{diag}$, is the diagonal state setting the off-diagonal elements to zero.

\begin{figure}
    \centering
    \includegraphics[width=0.48\textwidth]{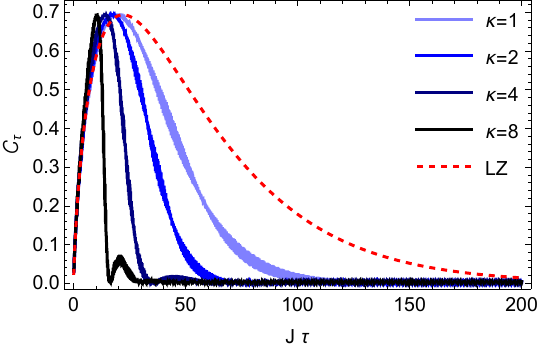}
    \caption{Relative entropy of coherence \eqref{eq:coh} at $t=\tau/2$ for a Gross-Pitaeveskii nonlinearity, $\tilde{\kappa}(z)=\kappa\, z$, and $\alpha=100$.}
    \label{fig:coherence_x2}
\end{figure}

In Fig.~\ref{fig:coherence_x2} we plot the relative entropy of coherence for the final state, $\mc{C}_\tau=\mc{C}(\rho_{\tau/2})$. Note that in Eq.~\eqref{eq:coh} we observe that $\mc{C}_\tau$ is a function of $\mc{P}_\tau$ only. Consequently, we note in Fig.~\ref{fig:coherence_x2} that the stronger the nonlinearity, the more coherences are suppressed for shorter driving times $\tau$. 

However, the relative entropy of coherence in Bloch representation does not give a conclusive picture, as the stationary states are not diagonal in energy, cf. also Ref.~\cite{Deffner2025QST}. Thus, it might still be possible that transitions between the effective energy states are suppressed at the expense of building up appreciable coherences between them. 

To investigate this, we also compute the Wigner-Yanese skew information \cite{Wigner1963PNAS,Luo2003PRL}, which is defined as
\begin{equation}
\mc{Y}(\rho,H)=-\frac{1}{2}\,\tr{\com{\sqrt{\rho}}{H}^2}\,,
\end{equation}
where we have chosen the Hamiltonian $H$ as conserved quantity. As is obvious from its definition, $\mc{Y}(\rho,H)$ measures how much a quantum states does not commute with a conserved quantity \cite{Takagi2019SR}, in our case $H$. This skew information has been extensively studied, as it finds application, e.g., in quantum speed limits \cite{Pires2016PRX}, metrology and uncertainty relations \cite{Muthuganesan2021PS,Chen2023JPA,Zhang2024EPJP}, and to quantify nonclassical correlations \cite{Chen2005PRA,Hong2024arXiv}.

For qubits, the Wigner-Yanase skew information can be further simplfied to read
\begin{equation}
\label{eq:WY}
\mc{Y}(\rho,H)=J^2 (1-x^2)-2 J \Delta x z+(1-z^2)\Delta^2\,,
\end{equation}
which in contrast to the relative entropy of coherence \eqref{eq:coh} also depends of the $x$-component of the Bloch vectors. This simple fact highlights why $\mc{Y}(\rho,H)$ is the more suitable quantifier of coherences for our purposes.

\begin{figure}
    \centering
    \includegraphics[width=0.48\textwidth]{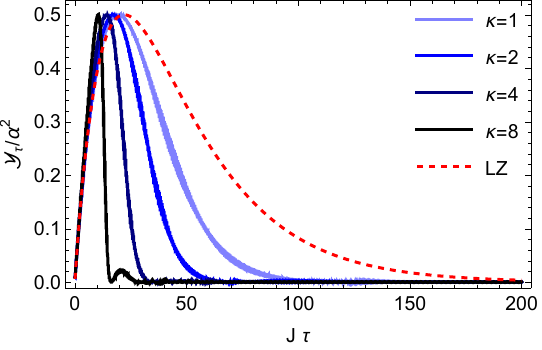}
    \caption{Wigner-Yanase skew information \eqref{eq:WY} at $t=\tau/2$ for a Gross-Pitaeveskii nonlinearity, $\tilde{\kappa}(z)=\kappa\, z$, and $\alpha=100$.}
    \label{fig:WY_x2}
\end{figure}

In Fig.~\ref{fig:WY_x2} we plot the resulting skew information, again for the Landau-Zener model driven through the ``avoided crossing'' in the presence of a Gross-Pitaeveskii nonlinearity. We observe, in complete analogy to the relative entropy of coherence, that the nonlinear term suppresses the build-up of coherences for shorter driving times $\tau$. This implies that for nonlinear dynamics, transitions as well as coherences are suppressed for faster processes, faster than in the adiabatic limit of the corresponding linear scenario. Therefore, we can now interpret the nonlinearity in Eq.~\eqref{eq:nonlin_sch} as an effective shortcut to adiabaticity for the linear version of the Landau-Zener model.

\section{Concluding remarks}

Quantum mechanics is a fundamentally linear theory. However, many practically relevant scenarios have been found, in which cooperative effects in complex quantum many body systems are well-described by effective, nonlinear Schr\"odinger equations. Such dynamics have many appealing properties ranging from unique computational capabilities \cite{Meyer2014PRA,Lacy2018QIP,Chiew2019QIP,Holmes2023PRA,Geller2023CTP,Geller2023AQT,Deiml2024arXiv} to superior thermodynamic properties \cite{Becker1946,Stankovic1973JPE,Chen2009,Feteira2009,Zhang2020MPLB,Deffner2025QST}. In the present work, we have shown that nonlinear dynamics can also be leveraged to facilitate effectively adiabatic dynamics for the corresponding linear system. The natural question arises whether the present findings are a peculiarity of the Landau-Zener model, or whether similar behavior can be found in more complex scenarios. 

In any case, nonlinear dynamics can do many things linear dynamics cannot. Our present results at least seem to suggest that further research into leveraging nonlinearities to control quantum dynamics is warranted.

\acknowledgements{Stimulating discussions with Emery Doucet and Moallison F. Cavalcante are gratefully acknowledged. This work was supported by the John Templeton Foundation under Grant No. 62422.}

\bibliographystyle{eplbib}
\bibliography{nonlinear}

\end{document}